\documentclass[twoside,slac_one]{revtex4}
\usepackage{graphicx}
\usepackage{fancyhdr}
\usepackage{amsmath} 
\usepackage{bm}
\usepackage{amsxtra}
\usepackage{amssymb}
\usepackage{amsthm}
\usepackage{latexsym}
\usepackage{lscape}

\begin{document}

\title{Study of Tau Leptons from Top Quark Pairs in ATLAS}

%

\author{A. McCarn, On Behalf of ATLAS}
\affiliation{Department of Physics, University of Illinois, Urbana-Champaign, IL, USA}

\begin{abstract}
Top quark pair decay events resulting in final states containing $\tau$ leptons present interesting channels in the search for physics beyond the Standard Model.  This document describes an analysis of $\tau$+jets in $t\bar{t}$ decays, for the case of a hadronically decaying  $\tau$ lepton.  The studied events are from 1.03~fb$^{-1}$ of proton-proton collision data at $\sqrt{s}=7$ TeV recorded by the ATLAS Collaboration at the LHC.  The goal of the analysis is to look for deviation from Standard Model expectations.  Exclusion limits are produced for the product of the branching ratios $BR(t\rightarrow H^+ b)$ and $BR(H^+ \rightarrow \tau \nu)$ with respect to the charged Higgs boson mass, and also on the $\tan(\beta)-M_{H^+}$ plane in the mh-max scenario of the MSSM.  

\end{abstract}

\maketitle

\thispagestyle{fancy}


\section{Introduction}
The decay of top quark pairs is a significant source of  $\tau$ leptons at the LHC.  Furthermore, top quark decays to  $\tau$ final states present an interesting channel in the search for new physics.  In particular, it is possible to calculate competitive limits on the exclusion of the light charged Higgs boson, a particle predicted by many non-minimal Higgs scenarios\cite{HiggsHunter,HiggsHunterErr}, such as Two-Higgs-Doublet-Models (2HDM)\cite{2HDM}.  One specific scenario considered is the mh-max scenario of the Minimal Supersymmetric Standard Model (MSSM), whose Higgs sector is described by a type-II-2HDM\cite{MSSMHiggs}.   

\begin{figure}[ht]
\centering
\includegraphics[width=80mm]{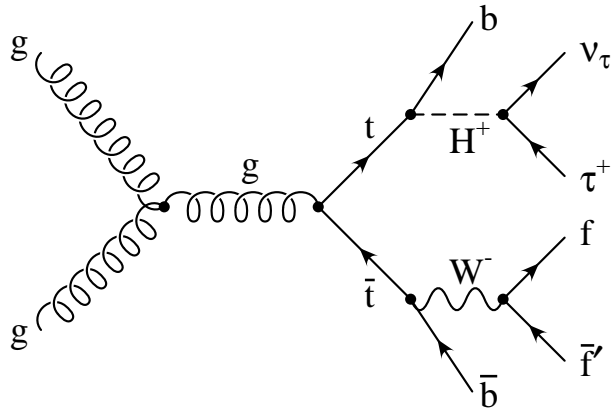}
\caption{Dominant production and decay of the light charged Higgs boson ($M_{H^+}$, smaller than the top quark mass, $m_t$) at the LHC .} 
\label{diagram}
\end{figure}

In the Standard Model, the top quarks in $t\bar{t}$ events are expected to decay exclusively into a bottom quark and $W$ boson.  $W$ bosons can then decay hadronically into light quarks, or leptonically into an $e$, $\mu$, or $\tau$ lepton and associated neutrino.   For charged Higgs boson masses, $M_{H^+}$, smaller than the top quark mass, $m_t$, the dominant production mode at the LHC for $H^+$ is the decay $t\rightarrow H^{+}b$ of one of the top quarks in $t\bar{t}$ events.  The charged Higgs boson decays via $H^{+}\rightarrow\tau\nu$ almost exclusively in many scenarios\cite{BR}. Therefore, the charged Higgs boson would be produced in the place of a $W$ boson for a fraction of top quark decays in $t\bar{t}$ events and would then decay exclusively to $\tau$ leptons, as shown in the diagram in Figure~\ref{diagram}.  Thus, if the charged Higgs boson exists, its preferential decay to $\tau$ leptons would cause an excess of events with $\tau$ leptons in the final state.

The $\tau$ lepton can decay either hadronically ($BR \approx 65\%$) or leptonically, into an $e$ or $\mu$ ($BR \approx 35\%$).  In the course of this study, only hadronically decaying $\tau$ leptons are considered, and any references to $\tau$ leptons refer only to only those decaying hadronically.  The channel under study is described by one $W$ (or charged Higgs) boson decaying into a hadronic $\tau$+$\nu$, and the other $W$ boson decaying hadronically.  This channel is referred to as ``$\tau$+jets".  Exclusion limits are calculated for the product of the branching ratios $BR(t\rightarrow H^{+} b)$ and $BR(H^{+} \rightarrow \tau \nu)$ with respect to the charged Higgs boson mass, and also on the $\tan(\beta)-M_{H^+}$ plane in the mh-max scenario of the MSSM. 

\section{Event Selections}
The $\tau$+jets final state is characterized by the presence of one hadronically decaying $\tau$ lepton, at least 4 jets, and missing transverse energy.  To select $\tau$+jets events, the following cuts are applied:

\begin{enumerate}
  \item Event preselection:
    \begin{enumerate}
      \item Data quality cuts.
      \item $E_{T}^{miss}$ and tau trigger.
      \item At least 4 jets with $p_T > 20$ GeV.
    \end{enumerate}
  \item Exactly one trigger matched $\tau$ jet with $ p_T^\tau >35$ GeV. 
  \item No identified $e$ or $\mu$.
  \item $E_{T}^{miss} >40 $ GeV.
  \item Reject events with large reconstructed $E_{T}^{miss}$ due to the limited resolution of the energy measurement with a cut on the significance of $E_{T}^{miss}$: $\frac{E_{T}^{miss}}{0.5 \cdot \sqrt{\sum E_T}} > 8 \;$  GeV$^{1/2}$. 
  \item At least one $b$-tagged jet.
  \item Topologies consistent with a top decay are identified by requiring that the $qqb$ candidate with the highest $p_T ^{qqb}$ value must satisfy $m(qqb) \in [120, 240]$ GeV.
\end{enumerate}

The final discriminating variable, which is used in limit setting, of the $\tau$+jets analysis is $M_T( \tau ,E_{T}^{miss})$, the transverse mass of the $\tau$ lepton and $E_{T}^{miss}$.  In the case of most of the Standard Model background ($t\bar{t}$ and $W$+jets) this quantity is related to the W boson mass, and in the case of the charged Higgs boson, this quantity is related to the charged Higgs mass.    

\section{Data Driven Fake Estimates}

An important aspect of the analysis is finding a way to estimate all of the backgrounds using data-driven methods. A method is developed for each significant source of background in the two analyses: jets being misidentified as $\tau$ leptons, electrons being misidentified as $\tau$ leptons, QCD multi-jet events, and events with true $\tau$ leptons in the final state.  This document focuses on the methods of estimating the contribution to background from jets being misidentified as $\tau$ leptons and QCD multi-jet events, though the other two methods are briefly outlined.

Hadronic $\tau$ leptons present a challenge in identification. They decay into either one or three charged tracks, a neutrino, and associated neutral particles.  The distinguishing features of hadronic $\tau$ decays in the detector are a low track multiplicity (1 or 3 charged tracks), and a narrow deposition of energy in the calorimeters.  These hadronic $\tau$ leptons look very similar to jets in the detector, and electrons are also sometimes misidentified as $\tau$ leptons with 1 associated charged track.   Understanding and estimating the misidentification of $\tau$ leptons is one of the major challenges of the analysis.

\subsection{Electron-to-$\boldsymbol{\tau}$ Misidentification}

Electrons can be misidentified as $\tau$ leptons with 1 associated charged track, but a dedicated electron veto algorithm greatly reduces the misidentification probability. To estimate the remaining background to each analysis from $e$-to-$\tau$ misidentification, a tag-and-probe method is used on $Z\rightarrow e^+e^-$ events from collision data.  The method used is identical to \cite{efakes}, but utilizes a larger dataset.  Instead of using the misidentification probability directly, a scale factor is derived from the difference between the misidentification probability in collision data and in $Z\rightarrow e^+e^-$ events in Monte Carlo.  This factor is then used to scale simulated events where the reconstructed $\tau$ originates from a true electron.  The contribution from events with electrons misidentified as $\tau$ leptons is less than $5\%$ of the total background. 

\subsection{Jet-to-$\boldsymbol{\tau}$ Misidentification}

Events with jets misidentified as $\tau$ leptons are the dominant source of backgrounds with misidentified $\tau$ leptons.  As a result,  the likelihood-based identification algorithm has been developed to distinguish $\tau$ leptons from light quark jets, the object most commonly misidentified as a $\tau$ lepton.  Even with the rejection power of the identification algorithm, light quark jets are often misidentified as $\tau$ leptons.  Less commonly, gluons or $b$-jets in $t\bar{t}$ and $W$+jets events can also be misidentified as $\tau$ leptons.  As the misidentification probability is higher for light quark jets than for gluons\cite{taufakes}, a data driven method based on the measurement of the light quark misidentification probability is taken as a conservative estimate of the contribution to the total background. 

The backgrounds from jets misidentified as $\tau$ leptons are estimated using misidentification probabilities taken from $\gamma$+jets events in collision data.  A $\gamma$+jets event is defined by passing a single $\gamma$ trigger, and by having at least one well-identified $\gamma$ and jet.  The leading $\gamma$ and jet are required to be back-to-back and balanced in $p_T$, and any additional jet is required to have less than $20\%$ of the $p_T$ of the lead $\gamma$.

\begin{figure*}[ht]
\centering
\includegraphics[width=67.5mm]{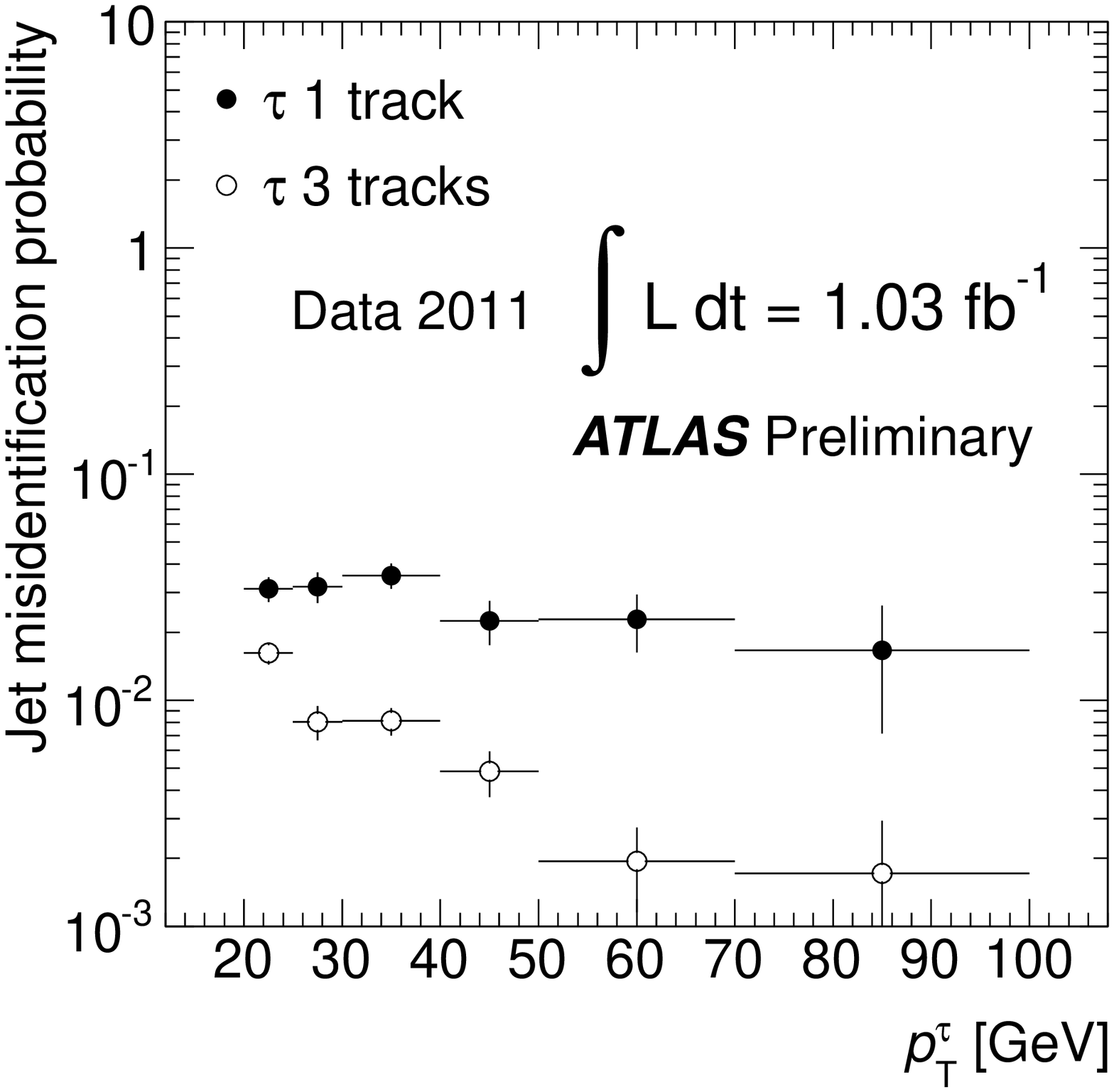}
\includegraphics[width=67.5mm]{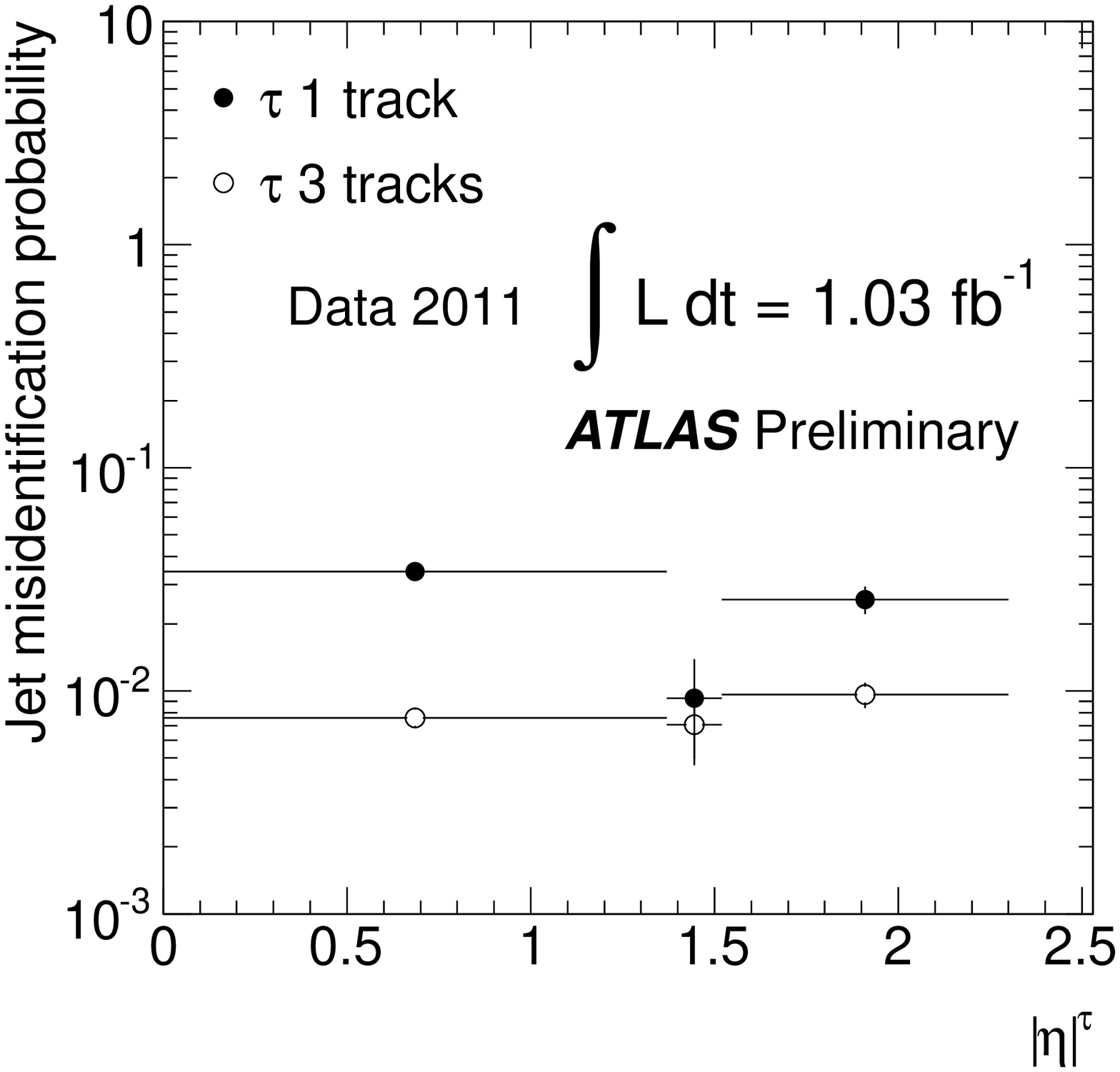}
\caption{Jet-to-$\tau$ Misidentification Probability. Misidentification probability evaluated for the tightest working point of the $\tau$ identification algorithm, corresponding to a $\tau$ efficiency of $30\%$. The probabilities pictured are for 1-prong and 3-prong $\tau$ leptons vs. $\tau$ $p_T$ (Left), and for 1-prong and 3-prong $\tau$s vs. $\tau$ $\eta$ (Right). } \label{Misid_Prob}
\end{figure*}

The $\gamma$ is the tag object, and the jet opposite is the probe object, through which the misidentification probability is calculated.  The denominator of this probability counts events which pass a loose $\tau$ selection.  This selection requires the $\tau$ candidate to have $p_T > 20$ GeV, to be seeded by a calorimeter-based algorithm, not be vetoed by the dedicated algorithm for rejecting electrons, and have exactly one or three associated charged tracks.  The numerator counts events which pass all of these requirements, in addition to a likelihood-based $\tau$ identification algorithm, which is optimized for accepting $\tau$ leptons and rejecting jets.

\begin{table}[ht]
\begin{center}
\caption{Background estimates from jet misidentification probability compared to the corresponding Monte Carlo prediction}
\begin{tabular}{|l|c|c|}
\hline
\multicolumn{3}{|c|}{\textbf{ATLAS} Preliminary } \\ 
\hline \textbf{Sample} & \textbf{Misid. Prob.} & \textbf{MC Prediction}
\\
\hline $t\bar{t}$ & $2.8 \pm 1.0$ & $3.8 \pm 0.6$ \\
\hline
\end{tabular}
\label{Misid_Prob}
\end{center}
\end{table}

The misidentification probability is dependent on both $p_T^{\tau}$ and $\eta^{\tau}$ and it is binned accordingly.  For $p_T$, the misidentification probability is split into six bins, with tighter binning at lower values and coarser with rising $p_T$.  For $\eta$, the probability is separated into the regions corresponding to the barrel region of the detector, the transition region between the barrel and the end cap, and the end cap.  The calculated misidentification probabilities are shown in Figure~\ref{Misid_Prob}, for one and three prong $\tau$ leptons with respect to $p_T$ and $\eta$.  In each case, the fake rate for a working point with $\tau$ efficiency of $30\%$ for the identification algorithm is displayed.  This tight working point in the $\tau$ identification algorithm is the one employed for the $\tau$+jets analysis.   

This misidentification probability is then applied to simulated events to predict the number of events arising from jets misidentified as $\tau$ leptons.  Table~\ref{Misid_Prob} shows the estimates based on the misidentification probability and the expectation from simulated events for the $\tau$+jets final states.  Events with jets misidentified as $\tau$ leptons make a very small contribution to the total background.

\subsection{QCD Multi-jet Estimate}

\begin{figure}[ht]
\centering
\includegraphics[width=80mm]{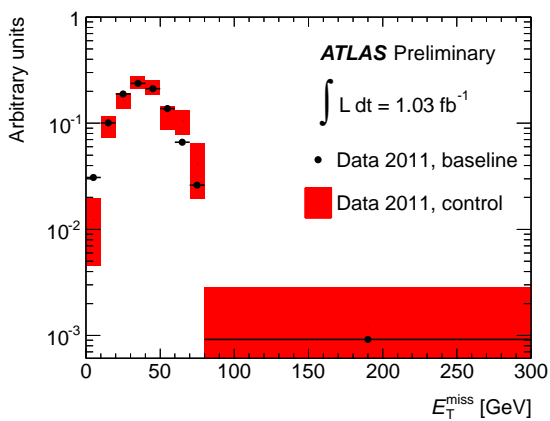}
\caption{Comparison of $E_{T}^{miss}$ distribution for the inverted and baseline $\tau$+jets selection early in the cut flow.  Inverted selection inverts $\tau$ identification cut and $b$-tag requirement.} \label{QCDtaujetsComp}
\end{figure}

The background from QCD multi-jet events is not well modeled in Monte Carlo, so it is estimated using a fitting method.  An inverted selection is defined that is orthogonal to the baseline selection.  The underlying assumption of the method is that the shape of the $E_{T}^{miss}$ distribution is the same for both the baseline and the inverted selection.  This is shown to be valid in the comparison of the $E_{T}^{miss}$ distributions for the baseline and inverted selections in ~Figure~\ref{QCDtaujetsComp}.  Differences in the two distributions are taken into account in the systematic uncertainties of the method.  

The orthogonal selection is defined by inverting the $\tau$ selection and the $b$-tag requirement.  The $\tau$ lepton must pass a loose identification cut while failing a tighter cut, and there must be no $b$-tagged jets in the event.  The $E_{T}^{miss}$ distributions for this case are shown for the inverted and baseline selections, after subtracting the non-QCD backgrounds estimated by simulated events, in Figure~\ref{QCDtaujetsComp}.  This comparison is made early in the cut flow, in order to have sufficient purity and statistics to make the comparison.  The two distributions match fairly closely, and the difference between the distributions is taken as a systematic uncertainty.

\begin{figure*}[ht]
\centering
\includegraphics[width=67.5mm]{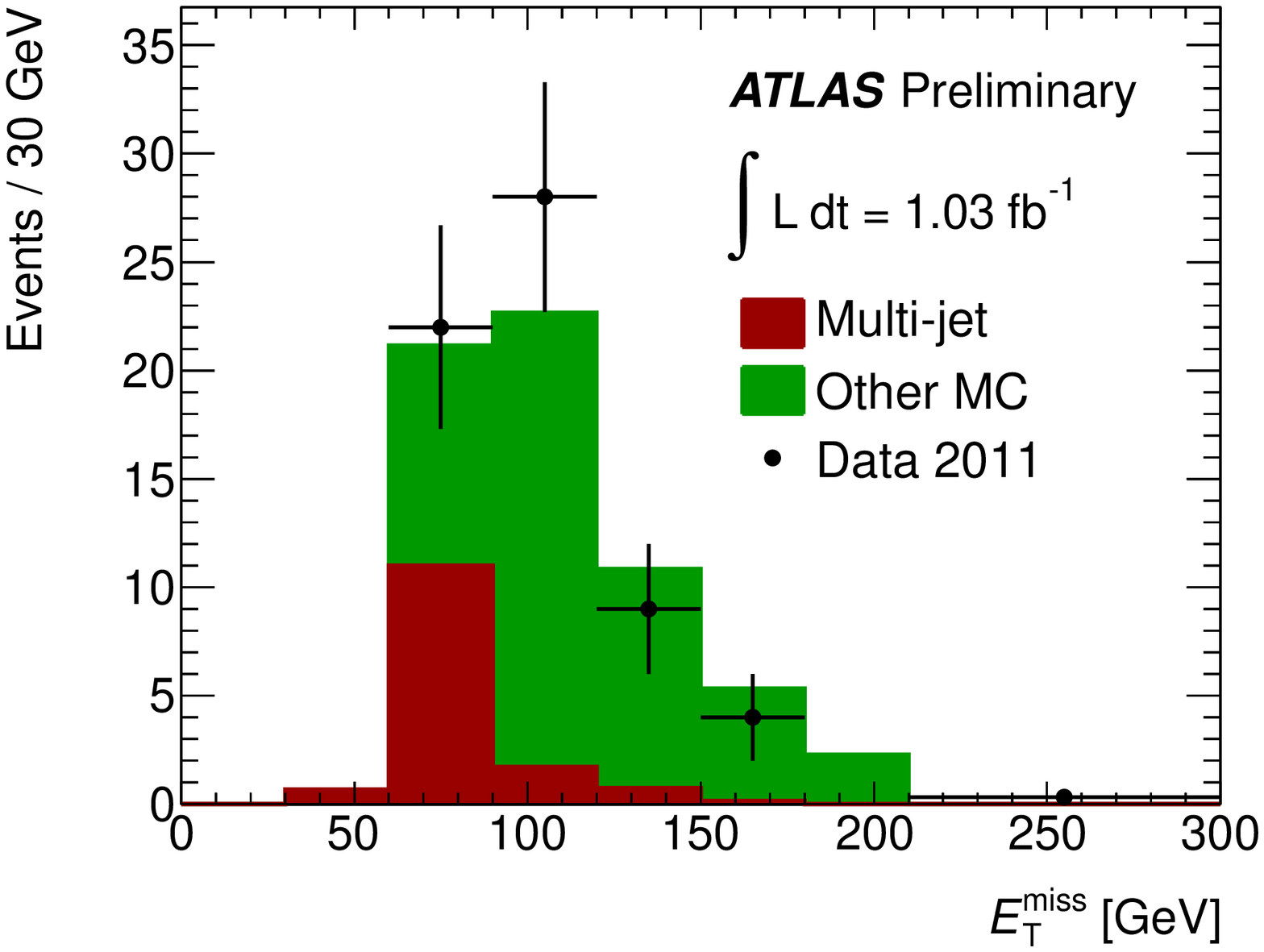}
\includegraphics[width=67.5mm]{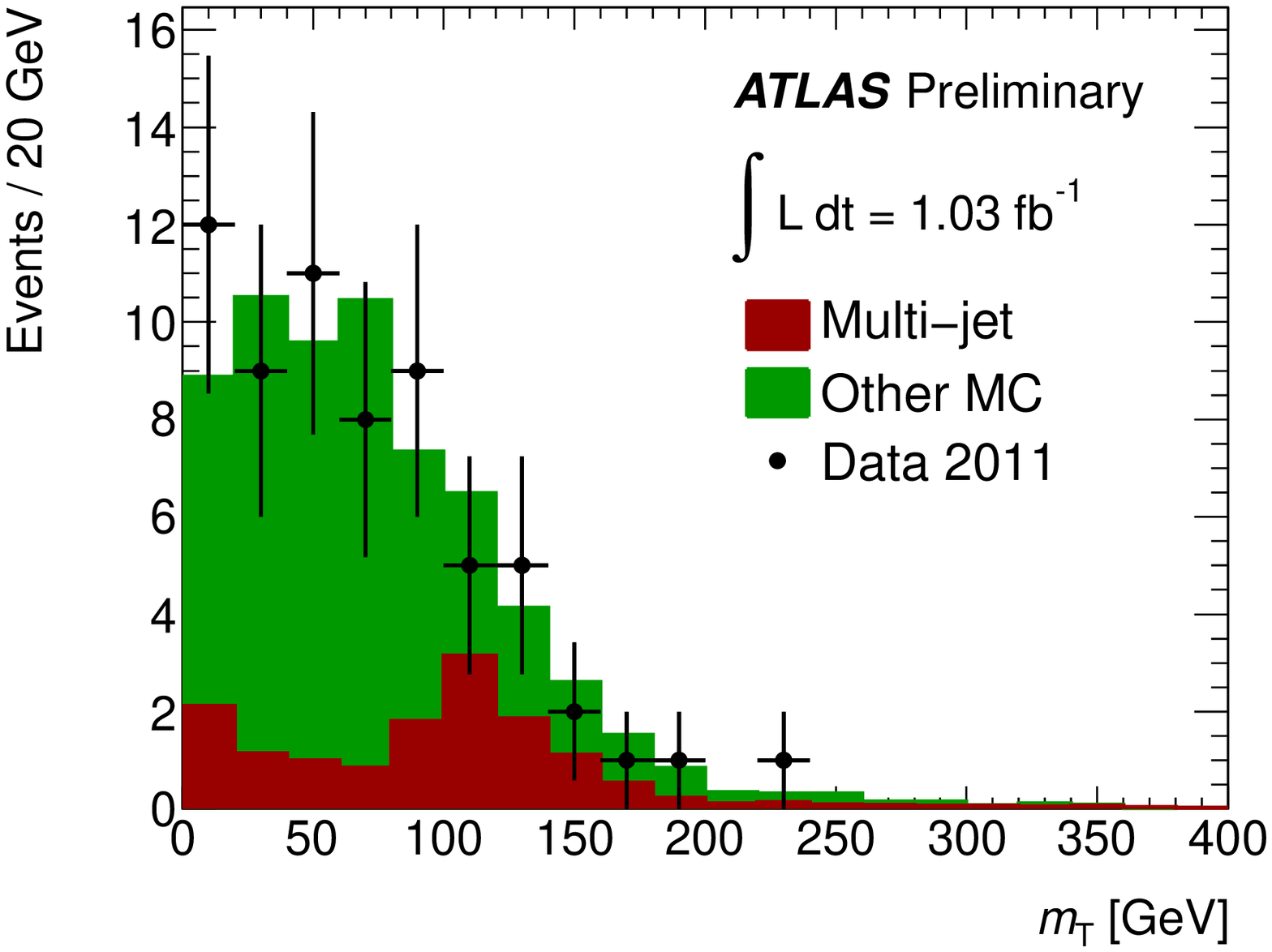}
\caption{Left: Results for the QCD fitting method for $\tau$+jets, $E_{T}^{miss}$ distribution. Right: Results applied to the $M_T(\tau,E_{T}^{miss})$ distribution.  } \label{QCDtaujets}
\end{figure*}

The inverted selection is taken as a model for the shape of the QCD background.  The two shapes used for the fit are the $E_{T}^{miss}$ distribution for this model and the $E_{T}^{miss}$ distribution of the non-QCD events from simulation.  The two free parameters of the fit are the overall normalization, which is fitted to the data, and the fraction of QCD multi-jet events.  The result of the fit is shown in the plot to the left in Figure~\ref{QCDtaujets}.  The plot to the right in Figure~\ref{QCDtaujets} shows the fit result applied to estimate the QCD contribution to the $M_T( \tau ,E_{T}^{miss})$ distribution.   $M_T( \tau ,E_{T}^{miss})$, the transverse mass of the $\tau$ lepton and $E_{T}^{miss}$, is the final discriminating variable for the $\tau$+jets analysis, and it is used in the limit setting process.

\subsection{Embedding Method}

Backgrounds with true $\tau$ leptons are estimated in a data-driven way using a method called embedding. Events with correctly identified $\tau$ leptons are a major background in the $\tau$+jets analysis. First, $\mu$+jets are selected with a high purity from data.  The muon signature (tracks, calorimeter deposition) is removed from the event, and a simulated $\tau$ with the same properties, but rescaled, is embedded in the event. Then, the full baseline selection is run over this embedded sample to estimate the total contribution to the background that results from events with true $\tau$ leptons in the final state.  This method has been validated in $\tau$+jets events using early ATLAS data\cite{embedding}.

\section{Results}

\begin{figure}[ht]
\centering
\includegraphics[width=80mm]{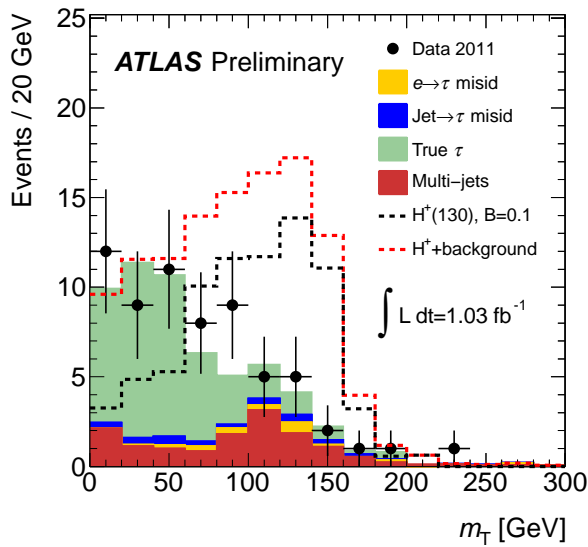}
\caption{Final plot of $M_T(\tau,E_{T}^{miss})$ for the $\tau$+jets analysis, using the data driven methods to estimate background contributions.  For reference, the charged Higgs boson signal and signal+background contribution for a charged Higgs boson mass of $130$ GeV and a $BR(t\rightarrow H^+ b) = 0.1$. } \label{final_taujets}
\end{figure}

\begin{figure}[ht]
\centering
\includegraphics[width=80mm]{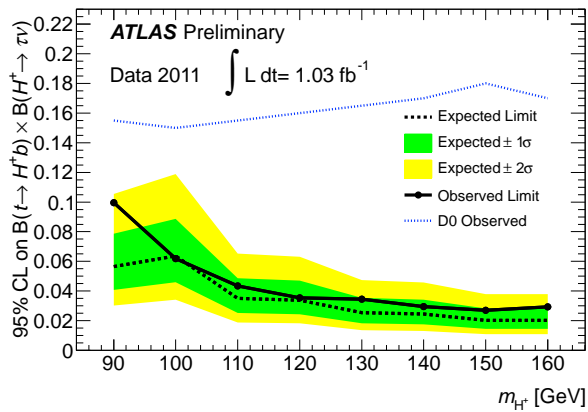}
\caption{Expected and observed $95\%$ CL exclusion limits for charged Higgs boson production from top quark decays as a function of $m_{H^+}$ in terms of $BR(t \rightarrow H^+ b)$ x $BR(H^+ \rightarrow \tau \nu)$. For comparison, the best limit provided by the Tevatron experiments is shown \cite{tevaLimit}.} 
\label{limit}
\end{figure}

\begin{figure}[ht]
\centering
\includegraphics[width=80mm]{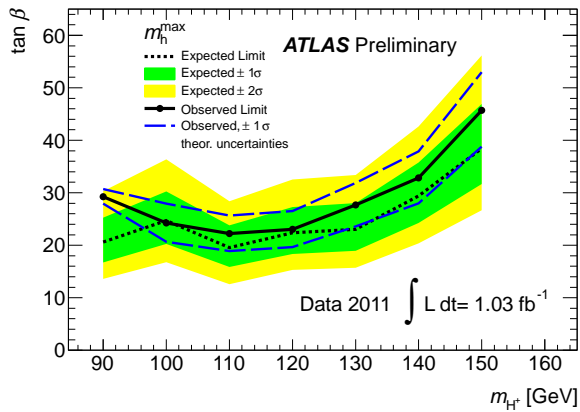}
\caption{Limit for charged Higgs boson production in top quark decays in the $M_{H^+}- \tan(\beta)$ plane.  Results are shown for the MSSM scenario mh-max} 
\label{mhmax}
\end{figure}

A final $M_T(\tau,E_{T}^{miss})$ distribution is shown in Figure~\ref{final_taujets}, with all background contributions taken from data-driven methods.  Also overlaid for reference are the signal and signal+background distributions for the case of a charged Higgs boson with a mass of  $130$ GeV and a $BR(t\rightarrow Hb) = 0.1$.  In the signal+background case, the background is the Standard Model background with altered branching ratios due to the hypothetical existence of a charged Higgs boson with $BR(t\rightarrow Hb) = 0.1$.  

Exclusion limits are set on the product of the branching ratios $BR(t\rightarrow H^+ b)$ and $BR(H^+ \rightarrow \tau \nu)$ with respect to the charged Higgs boson mass, and also on the $\tan(\beta)-M_{H^+}$ plane in the mh-max scenario of the MSSM, by applying the CL$_s$ procedure\cite{CLS1,CLS2}. A profile likelihood ratio \cite{llh} is used with the $M_T(\tau,E_{T}^{miss})$ as the discriminating variable.   Shown in Figure~\ref{limit}, values of the product of branching ratios, $BR(t\rightarrow H^+ b)$ and $BR(H^+ \rightarrow \tau \nu)$, larger than $0.03 \-- 0.10$ have been excluded in the $H^+$ mass range $90 \-- 160$ GeV, significantly extending limits from other experiments. Results interpreted in the context of the mh-max scenario of the MSSM are shown in Figure~\ref{mhmax}.  Values of $\tan(\beta) $ above $22 \-- 30$ (depending on $m_{H^+}$) can be excluded in the mass range $90$ GeV $< m_{H^+} < 140$ GeV.

\begin{acknowledgments}
These analyses have been performed in the context of the Higgs to Complex States Working Group in the ATLAS Collaboration, and also rely on methods and results from the Tau Working Group and the Top Working Group.  The data driven methods for predicting the backgrounds are described in \cite{MoriondNote}. 
\end{acknowledgments}

\bigskip 

\bibliographystyle{plain}
\bibliography{myrefs}


\end{document}